# Dielectric Anomalies in Crystalline Ice: Indirect evidence of the Existence of a Liquid-liquid Critical Point in $H_2O$

Fei Yen[1,2*], Zhenhua Chi[1,2], Adam Berlie[1†], Xiaodi Liu[1], Alexander F. Goncharov[1,3]

[1]Key Laboratory for Materials Physics, Institute of Solid State Physics, Hefei Institutes of Physical Science, Chinese Academy of Sciences, Hefei 230031, P. R. China
[2]High Magnetic Field Laboratory, Hefei Institutes of Physical Science, Chinese Academy of Sciences, Hefei 230031, P. R. China
[3]Geophysical Laboratory, Carnegie Institution of Washington, 5251 Broad Branch Road, NW, Washington D.C., 20015, USA

[*]**Correspondence:** fyen18@hotmail.com
[†]**Present address:** ISIS Neutron and Muon Source, STFC Rutherford Appleton Laboratory, Didcot, Oxfordshire OX11 0QX United Kingdom

**Abstract:** The phase diagram of $H_2O$ is extremely complex; in particular, it is believed that a second critical point exists deep below the supercooled water (SCW) region where two liquids of different densities coexist. The problem however, is that SCW freezes at temperatures just above this hypothesized liquid-liquid critical point (LLCP) so direct experimental verification of its existence has yet to be realized. Here, we report two anomalies in the complex dielectric constant during warming in the form of a peak anomaly near $T_p$=203 K and a sharp minimum near $T_m$=212 K from ice samples prepared from SCW under hydrostatic pressures up to 760 MPa. The same features were observed about 4 K higher in heavy ice. $T_p$ is believed to be associated to the nucleation process of metastable cubic ice I$c$ and $T_m$ the transitioning of ice I$c$ to either ices I$h$ or II depending on pressure. Given that $T_p$ and $T_m$ are nearly isothermal and present up to at least 620 MPa and ending as a critical point near 33-50 MPa, it is deduced that two types of SCW with different density concentrations exists which affects the surface energy of ice I$c$ nuclei in the "no man's land" region of the phase diagram. Our results are consistent with the LLCP theory and suggest that a metastable critical point exists in the region of 33–50 MPa and $T_c \geq 212$ K.

**TOC Graphic:**

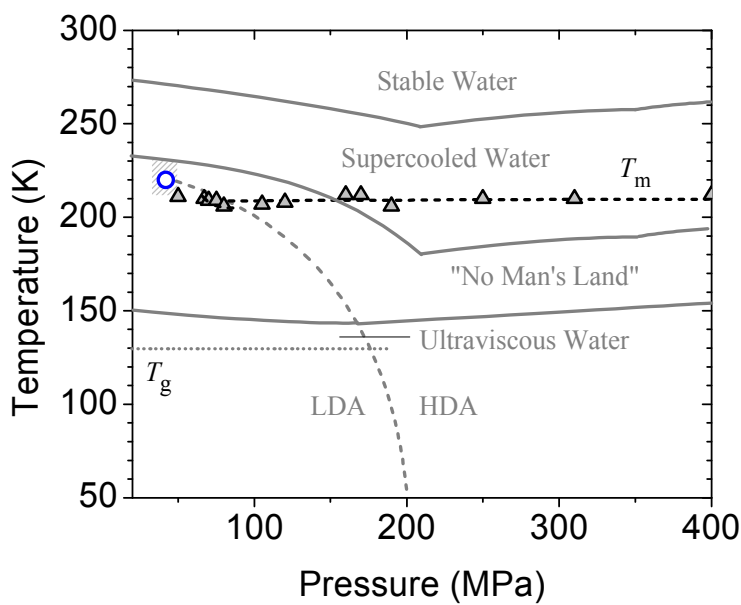

$H_2O$ ice is one of the most abundant solids in our universe with vast quantities existing on Earth and within our solar system. As such, many researchers are interested in knowing the structures, processes and patterns that these different ice phases exhibit.[1] Depending on the temperature and pressure, the oxygen atoms in ice form different types of crystallographic configurations (Fig. 1) such as hexagonal (ice I),[2] rhombohedral (ice II),[3] tetragonal (ices III, IX and VI),[4-6] monoclinic (ices V and XIII),[7,8] other interpenetrating frameworks (ices VII and VIII),[9,10] and even cubic (ice X) at ultrahigh pressures.[11,12] The hydrogen atoms (designated as the protons) on the other hand, are more flimsy and are only restricted by the ice rules.[13] Namely, a) each proton is linked to two oxygen atoms, one via an intramolecular covalent bond and another via an intermolecular van der Waals hydrogen bond, and b) each oxygen atom is linked to four protons, two via covalent bonding and two via hydrogen bonding. At low temperatures the hydrogen atoms also form a crystallographic pattern like the oxygen atoms and the respective phases are considered to be "proton ordered". Much of the complexities in ice arise from the intra- and intermolecular bonding of the hydrogen atoms.[14]

Despite the simplicity of the $H_2O$ molecule, over a dozen different crystallographic phases (ices I$h$, Ic,[15] II–XV), at least three amorphous phases,[16,17] a supercooled region,[18] and a 'no man's land' occupy the phase diagram.[19] Apart from the well established critical point where water and ice coexist, a second critical point is believed to reside near 210 K and 100 MPa (Ref. [20]) separating two types of liquid structures composed of different concentrations of a low density liquid and a high density liquid.[21,22] The problem however, is that supercooled water (SCW) freezes at a temperature just above this hypothesized liquid-liquid critical point (LLCP) which makes it nearly impossible to access experimentally. As such, much work has been devoted on studying the metastable extensions of several phase boundary lines that extend toward the expected location of the LLCP,[23,24,25] as well as *nano-*,[26] and *micro*-sized[27] samples to confirm the existence (or nonexistence) of the LLCP. As Stanley *et al*. propose,[28] other indirect approaches are needed to confirm its existence which we attempt in this work. Verification of the existence of the LLCP will bring consistency to the phase behavior of metastable water and ice[29] and further understand other systems that also possess a liquid-liquid phase boundary such as the cases in silicon,[30,31] phosphorous[32] and cerium.[33]

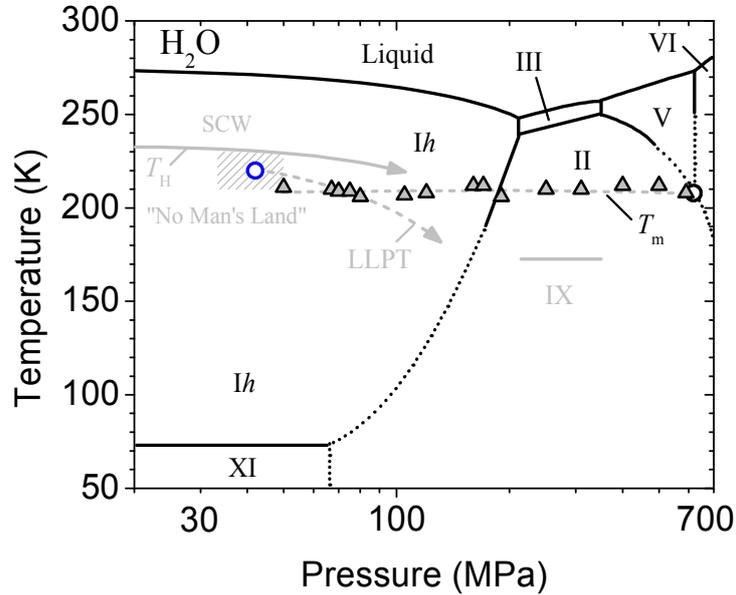

Fig. 1: Low pressure low temperature phase diagram of $H_2O$. All dotted lines and the triple point (black circle) near 620 MPa and 208 K are based on guesswork. All phases and phase lines in grey are metastable. $T_m$ (this work) represents the transition of metastable ice I$c$ into ices I$h$ or II depending on pressure exhibited by a sharp minimum anomaly in the real and imaginary parts of the dielectric constant. $T_m$ ends as a critical point somewhere in the region of $33<P_c<50$ MPa and 210 K where we conjecture that this is the same critical pressure where a liquid-liquid phase transition (LLPT) line separating two types of liquid structures also end as suggested in Ref. [20]. $T_H$

is the homogeneous nucleation curve separating supercooled water (SCW) and "No Man's Land". Blue circle is plotted at 42 MPa and 220 K (average of 210 K and the value of $T_H$ at 42 MPa).

The dielectric constant is a macroscopic quantity that is proportional to the dipole moment distribution at the atomic level. Hence, any structural transition is readily reflected in the dielectric constant because the bond angles and lattice parameters which constitute the charge distributions undergo changes. During the proton (dis)ordering process, the imaginary part of the dielectric constant is particular sensitive since there are movement of charges involved. For this reason, we carried out high precision measurements of the complex dielectric constant on samples of $H_2O$ prepared from SCW as a function of temperature up to 760 MPa. We first present a detailed analysis of an isobar at 310 MPa as an example to understand the different transformations that ice undergoes under pressure when subjected to cooling followed by warming. This will help appreciate the two new anomalies we identify in the form of a broadened peak near $T_p$=203 K followed by a sharp minimum near $T_m$=212 K during warming in the pressure range of 50–620 MPa. We attribute $T_p$ to be associated to the nucleation process of small amounts of cubic ice (ice I$c$) and $T_m$ its subsequent transition to either ice I$h$ or II in the "no man's land" region of the metastable phase diagram of water. $T_m$ is also intrinsic in $D_2O$ (D=deuterium) and found to completely vanish at pressures below 33 MPa suggesting that an LLCP indeed exists in 'no man's land'.

Figures 2a and 2b show the real $\varepsilon'(T)$ and imaginary parts $\varepsilon''(T)$ of the dielectric constant as a function of temperature at 310 MPa, respectively. At $T_{L\_III+SCW}$=252 K during cooling, part of the sample transformed into ice III represented by a shoulder type discontinuity in $\varepsilon'(T)$ and $\varepsilon''(T)$ coinciding exactly with the H$_2$O liquidus line.[34] The residual liquid that did not transform remained as supercooled water (SCW) until $T_{\text{III+SCW\_III}}$=232 K where heterogeneous nucleation of ice III occurred marked by a sharp drop in both $\varepsilon'(T)$ and $\varepsilon''(T)$. This process is one method how crystalline ice III is formed.[34] In the range of 211>$T$>164 K, $\varepsilon'(T)$ decreased by over half its value while $\varepsilon''(T)$ exhibited a minimum at 210 K and a maximum at 203 K. The dielectric constant within an ice phase usually does not change by much so the origin of such a drastic change can be presumed to stem from proton ordering dynamics. Indeed, Whalley *et al.* concluded that continuous proton ordering occurs in ice III from 208 K to 163 K to end up forming ice IX.[5]

It was later confirmed by neutron diffraction on D$_2$O (Refs. [4,35]) that only 96% of the deuterons ordered at ~77 K. According to DFT calculations from Ref. [36], the protons should undergo a first-order phase transition to become a fully ordered antiferroelectric ground state at $T_{\text{IX'}}$=126 K. Unfortunately, Whalley *et al.*,[5] only presented data down to 158 K and the neutron diffraction studies were not carried out *in situ* nor under hydrostatic pressure conditions. In our case, a change of slope discontinuity was observed in our data in $d\varepsilon''(T)/dT$ (inset of Fig. 2b) which coincides exactly with $T_{\text{IX'}}$.

Upon warming, both $\varepsilon'(T)$ and $\varepsilon''(T)$ took a different path from their cooling curves starting from $T$>164 K which coincides with the fact that ice IX transforms into ice II upon warming.[5] At $T_{\text{II\_III}}$=245 K, a sharp step-up indicates the phase transition from ice II to ice III, also in good agreement with existing literature.[34] The region below $T_{III+SCW\_III}$=232 K of the cooling curve and above $T_{II\_III}$=245 K of the warming curve in both ε'(T) and ε"(T) appear to be parts of the same function which is not surprising because in both of these two regions, the system was in the same ice III phase.

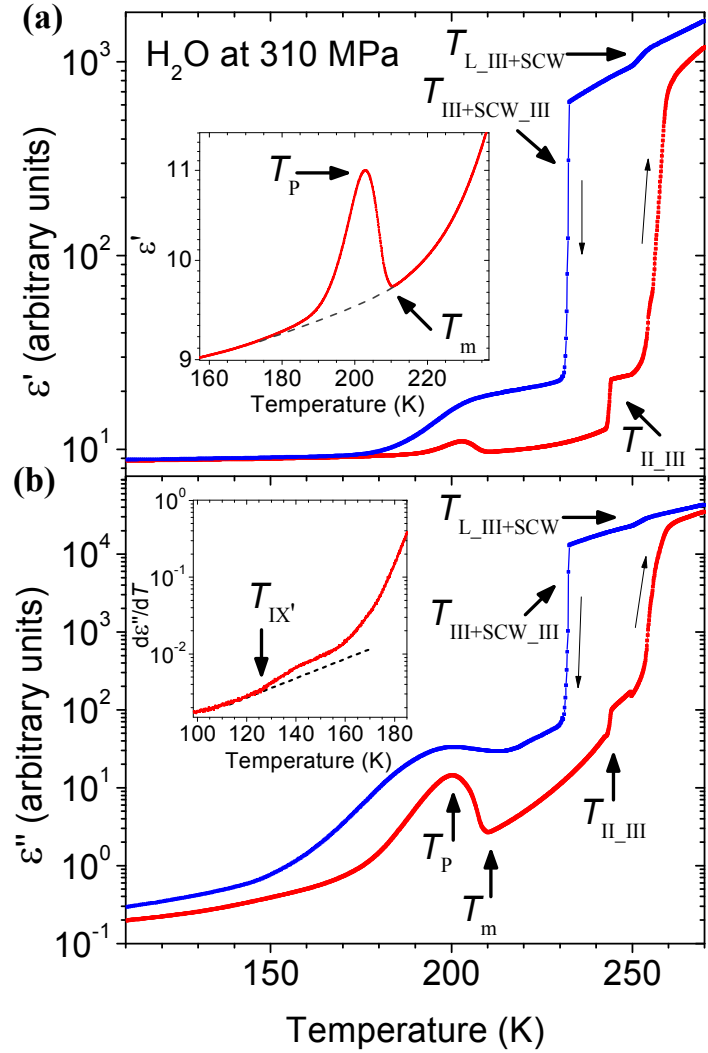

Fig. 2: Dielectric constant of $H_2O$ at 310 MPa. (a) Real ε'(T) and (b) imaginary ε"(T) parts of the dielectric constant. At $T_{L\_III+SCW}$=252 K, part of the sample transformed into ice III while the rest remained as supercooled water (SCW). At $T_{III+SCW\_III}$=232 K, the remaining SCW nucleated into ice III. At $T_{II\_III}$=245 K, the system transformed back to ice III. Two new anomalies are identified upon warming: a peak at $T_p$=203 K associated to the nucleation process of ice Ic; and a sharp minimum at $T_m$=210 K due to the transitioning of ice Ic into ice II. Inset of (b): An anomaly in the form of a linear to nonlinear change in dε"(T)/dT at $T_{IX'}$=126 K suggests that full ordering of the protons occurred at $T<T_{IX'}$ according to Ref. [36].

The new features we observe are a peak anomaly during warming in ε'(T) and

ε"(T) at $T_p$=203 K followed by a sharp minimum at $T_m$=210 K (inset of Fig. 2a and Fig. 2b). To understand more about the nature of $T_m$, in one of the scans at 190 MPa, the temperature was lowered immediately after the sample passed through $T_m$ (curve 3 in Fig. 3a). Interestingly, both $T_p$ and $T_m$ were not observed during cooling which classifies these two features as metastable. The temperature was lowered all the way down to 77 K followed by warming back to room temperature shown as curve 4. Surprisingly, both $T_p$ and $T_m$ were also not observed during this second warming run hinting that these two anomalies may have been a consequence of the cooling process. It should be noted that a third cool down starting from room temperature and subsequently warmed up yielded the same results as curves 1 and 2.

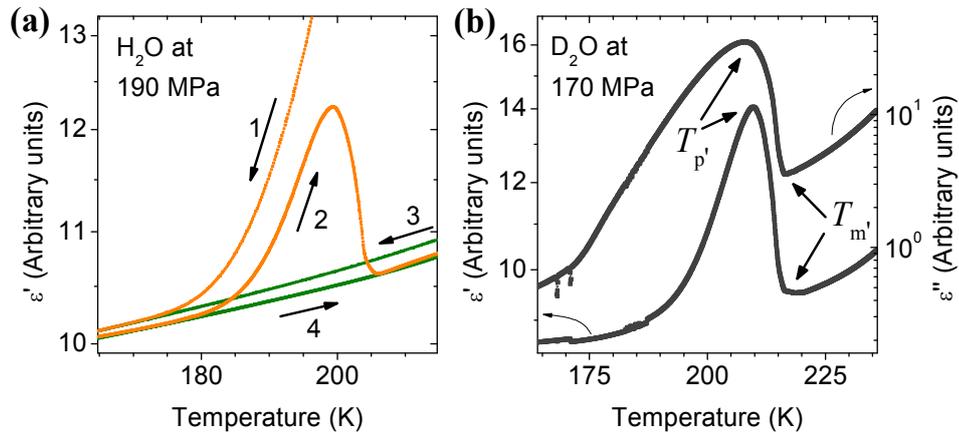

Fig. 3: (a) ε'(T) curves subjected to various initial conditions: curve 1 is the initial cool-down from 300 K to 77 K, curve 2 is warming from 77 to 220 K, curve 3 is cooling from 220 K back to 77 K and curve 4 is warming from 77 to 300 K. (b) is ε'(T) and ε"(T) in the case for $D_2O$ during warming at 170 MPa where the two anomalies were also observed at $T_{p'}$=208 K and $T_{m'}$=216 K.

Fig. 3b shows ε'(T) and ε"(T) for $D_2O$ at 170 MPa where $T_{p'}$=208 K and $T_{m'}$=216 K. The phase boundaries of $D_2O$ are usually about 4 degrees higher than $H_2O$ (Ref. [37]) so both sets of our results are in good agreement with each other.

Figures 4a-d show different warming curves of ε'(T) and ε"(T) from 33 to 720 MPa. In the low pressure region, the maximum in $T_p$ and minimum in $T_m$ were still

clearly evident in the 70 MPa curve. For the 67 and 50 MPa curves, the maxima and minima features were diminished and evolved into a step anomaly. Finally at 33 MPa, $T_p$ and $T_m$ were undetected indicating the existence of a metastable critical point. At the high pressure end, $T_m$ appears to intercept with the ice II/V/VI triple point estimated to reside near 620 MPa and 208 K.[38] It is not known if $T_m$ and $T_p$ completely vanish at $P>620$ MPa as they may have been masked by the dielectric constant of ice VI of which is the largest of all the ices.

During the freezing process multiple nucleation sites appear so grain boundaries will ultimately be present as the formation of one entire single crystalline piece is not probable. In between the crystalline regions pockets of topologically disordered material exist. We do not know the size of the pockets formed as the kinetics of ice nucleation in "no man's land" is still not well understood even at ambient pressure conditions.[39,40] However, it is reasonable to presume that these pockets and interfaces are no more than a few nanometers in size. It is now well known that ice I$c$ is the most favorable type of ice to form in confined volumes and interfaces that are, respectively, 15 and 10 nm or smaller in the 160-220 K range.[41,42] For instance, the type of ice that forms from supercooled water is a stacking type of ice I$c$ and ice I$h$ with a ratio of about 2 to 1.[43]

Interestingly, ice I$c$ has also been reported to form from amorphous ice[44] and high pressure ices;[45] its properties and degree of reconstruction of its stacking faults depend on the initial state of formation and surrounding temperature[1] (meaning that different variants of ice I$c$ can exist). Specifically, ices II, III and V transform to ice I$c$ upon warming at ambient pressure in the ranges of 168–178 K, 148–158 K and 151–155 K, respectively.[45] Upon further warming, ice I$c$ transforms into ice I$h$ at 180-228 K.[46,47] The reason why metastable ice I$c$ forms at low temperature is because the surface energy of its nuclei is lower than ice I$h$.[48] However, at higher temperature, nucleation of ice I$h$ is favored because the surface entropy of its nuclei is larger than that of ice I$c$.[49] Thus, we attribute $T_p$ to be associated to the nucleation process of ice I$c$ of which its nuclei most likely formed during the cooling process which explains why a minimum in $\varepsilon''(T)$ exists near 203 K in the cooling curves (Fig. 2b). This is key toward understanding the kinetics of ice nucleation in "no man's land" as we are merely in the beginning stages of understanding its details in SCW.[50] Hence, $T_p$ marks the point where the sample comprised the highest amount of ice I$c$; and $T_m$ the abrupt transformation of ice I$c$ to either ice I$h$ or ice II depending on the pressure environment. Note that ice I$c$ is metastable in the phase spaces of ices I$h$ and II so once ice I$c$ transitions into ices I$h$ or II upon warming, it can no longer transition back; this is consistent with the metastable behavior of $T_m$ and $T_p$ (Fig. 3a). For instance, in Fig. 2a, ice I$c$ nucleated into ice II at $T_m$ so in the region of $T_m<T<245$ K the entire sample was in the ice II phase.

It should also be noted that ice I$c$ has only been reported to form from quenching[42] with the slowest cooling rate being 10 K-min$^{-1}$ (Ref. [47]) at ambient conditions. To our knowledge, no work on ice I$c$ under pressure has ever been reported. In our work, our cooling rates have been at most 3 K-min$^{-1}$ so formation of ice I$c$ is not expected near ambient pressure conditions which is consistent with the fact that $T_p$ and $T_m$ are absent up to 33 MPa. Only starting from 50 MPa do $T_p$ and $T_m$ appear. Pressure somehow favors the formation of ice I$c$ when cooled slowly either by lowering the surface energy of its nuclei or by further reducing the volumes of the pockets of trapped disordered material. Starting from 50 MPa the $T_p$ and $T_m$ features grow with pressure; however, their critical temperatures remain nearly unchanged for a large range of pressure which suggests that there are two types of liquid structures that affect the surface energy of ice I$c$ nuclei differently. If we take into account that the LLCP is situated in the vicinity near 100 MPa and 220 K as originally suggested in Ref. [20], and that a liquid-liquid phase transition (LLPT) line extends out of the LLCP toward lower temperature and higher pressure separating a low pressure and a high pressure liquid structure (Fig. 1), an indirect correlation can be drawn between the LLCP and the critical point obtained in this work.

Below the homogenous nucleation curve $T_H$, it is not energetically favorable for $H_2O$ to exist in liquid form. The properties of ice I$c$ are unique in that it is more fluid-like than all of the other crystalline phases as its stacking faults can be reconfigured depending on the surrounding environment.[1] Hence, out of all the possible ice phases, ice I$c$ appears to be the only phase that would be susceptible enough to changes affecting its nucleation process should there be two different types of liquid structures residing below $T_H$. Based on the correlation drawn between the LLCP and the critical point identified in this work, we suggest that the critical pressure of the LLCP resides in the pressure range of $33<P_c<50$ MPa. As for the critical temperature, we suggest that $T_H>T_c \geq T_m$ since $T_m$ is only representative of the transition temperature of ice I$c$ to ices I$h$ or II. The deduced pressure and temperature ranges are drawn as a grey box in Fig. 1 along with its average values marked as a blue circle at 42 MPa and 220 K. The obtained values for $P_c$ and $T_c$ are in agreement with many of the values extracted from indirect experiments[23-25,51] and predicted by recent simulations models.[52-55]

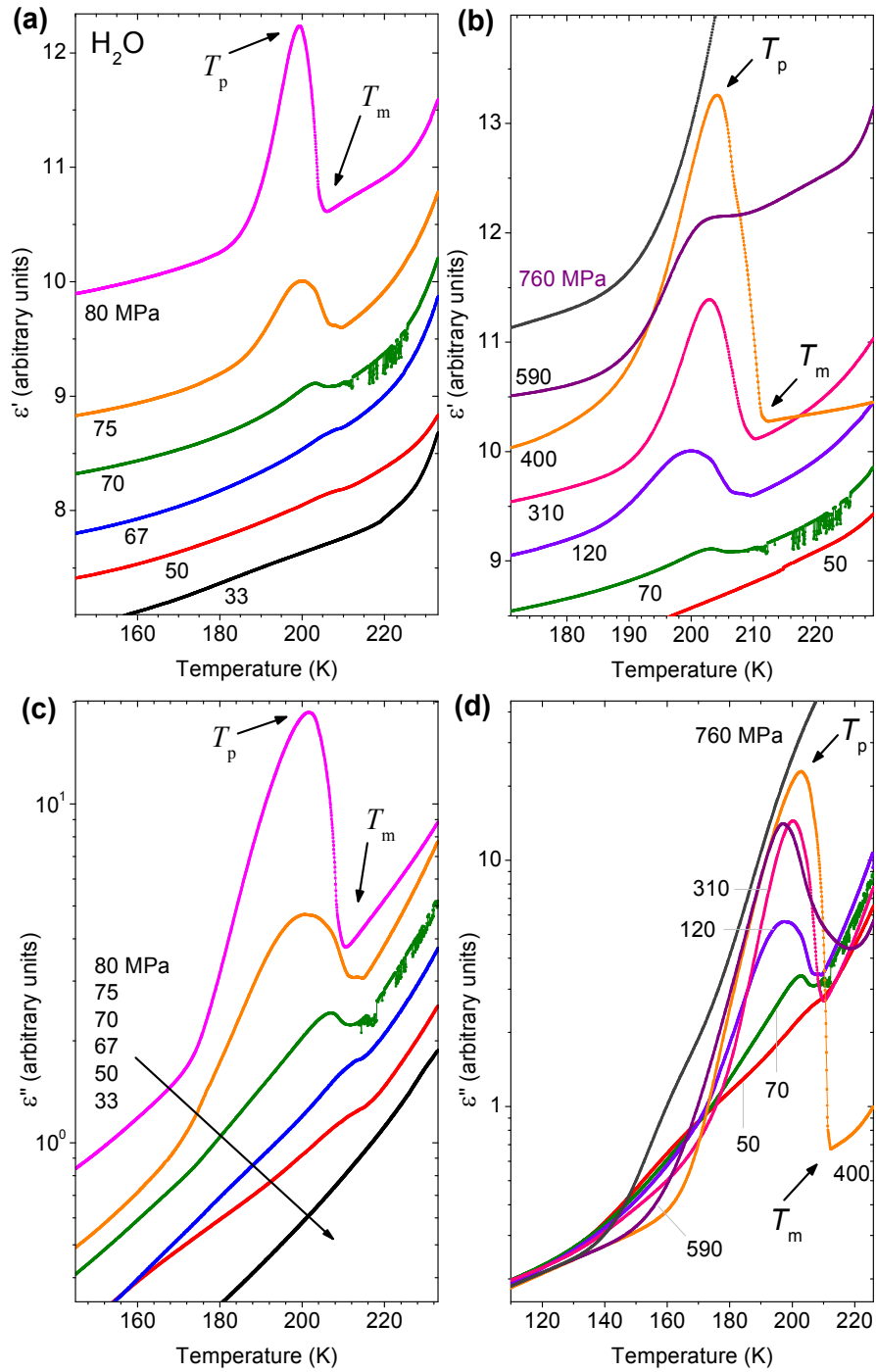

Fig. 4: Dependence of $T_p$ and $T_m$ in $\varepsilon'(T)$ and $\varepsilon''(T)$ at different hydrostatic pressure. The 110 and 400 MPa isobars were from a different sample. The $T_m$ values at different pressure are plotted in Fig. 1.

This work was made possible in part via the support of the National Science Foundation of China grant numbers 11374307 and 51372249, and the Director's Grants of the Hefei Institutes of Physical Science, Chinese Academy of Sciences grant number YZJJ201313.

**Methods:**

The samples were prepared by pressurizing deionized and degassed, liquid $H_2O$ (Milli-Q Direct 8) to the desired pressure at room temperature with a BeCu clamp cell then cooled to 77 K at ~2-3 K·min$^{-1}$ via a customized gas exchange cryostat. For the $D_2O$ experiments, 99.9% isotopic purity from Sigma Aldrich) was used and the same procedure was followed. The real and imaginary parts of the dielectric constant were obtained by measuring the capacitance and loss tangent, respectively, of a pair of Pt electrodes in the form of parallel plates at 1 kHz with an Andeen Hagerleen (AH2500A) capacitance bridge. The electrodes were dipped inside a Teflon capsule filled with liquid $H_2O$ so the sample itself was also the pressure medium. More details can be found elsewhere.[56] Force was applied only when the sample was in liquid form so hydrostatic conditions were maintained upon ice nucleation.